\documentclass[twocolumn]{aastex7}

\usepackage{graphicx}
\usepackage{amsmath,bm}
\usepackage[normalem]{ulem}
\usepackage{url}
\usepackage{multirow}
\usepackage{appendix}

\def \aap  {A\&A}

\def \apjs  {ApJS}
\def \apjl  {ApJL}

\renewcommand{\deg}{^{\circ}}

\newcommand{\temp}[1]{#1}

\begin{document}

\title{The role of three-dimensional effects on ion injection and acceleration in perpendicular shocks}

\author[0000-0002-1879-457X]{Luca Orusa}
\affiliation{Department of Astrophysical Sciences, Princeton University, Princeton, NJ 08544, USA}
\email[show]{luca.orusa@princeton.edu}  
\affiliation{Department of Astronomy and Columbia Astrophysics Laboratory, Columbia University, New York, NY 10027, USA}
\author[0000-0003-0939-8775]{Damiano Caprioli}
\affiliation{Department of Astronomy \& Astrophysics, University of Chicago, Chicago, IL 60637, USA}
\email{luca.orusa@princedon.edu} 
\affiliation{Enrico Fermi Institute, The University of Chicago, Chicago, IL 60637, USA}
\author[0000-0002-1227-2754]{Lorenzo Sironi}
\affiliation{Department of Astronomy and Columbia Astrophysics Laboratory, Columbia University, New York, NY 10027, USA}
\affiliation{Center for Computational Astrophysics, Flatiron Institute, 162 5th avenue, New York, NY, 10010, USA}
\email{luca.orusa@princedon.edu} 
\author[0000-0001-9179-9054]{Anatoly Spitkovsky}
\affiliation{Department of Astrophysical Sciences, Princeton University, Princeton, NJ 08544, USA}
\email{luca.orusa@princedon.edu} 

\begin{abstract}
Understanding the conditions that enable particle acceleration at non-relativistic collisionless shocks is essential  to unveil the origin of cosmic rays. We employ 2D and 3D hybrid simulations (with kinetic ions and fluid electrons) to explore particle acceleration and magnetic field amplification in non-relativistic perpendicular shocks, focusing on the role of shock drift acceleration and its dependence on the shock Mach number. 
We perform an analysis of the ion injection process and demonstrate why efficient acceleration is only observed in 3D. In particular, we show that ion injection critically depends on the "porosity" of the magnetic turbulence in the downstream region near the shock, a property describing how easily the post-shock region allows particles to traverse it and return upstream without being trapped. This effect can only be properly captured in 3D.
Additionally, we explore the impact of numerical resolution on ion energization, highlighting how resolving small-scale turbulence -- on scales below the thermal ion gyroradius -- is essential for accurately modeling particle injection. 
Overall, our results emphasize the necessity of high-resolution 3D simulations to capture the fundamental microphysics driving particle acceleration at perpendicular shocks.

\end{abstract}



\section{Introduction}

Particle acceleration in astrophysical non-relativistic shocks is a central topic in Astroparticle Physics, at the confluence of high-energy astrophysics and plasma physics \citep{krymskii77, axford+77p, bell78a,blandford+78}. 
In astrophysical environments shocks are typically collisionless, meaning that the conversion of energy from bulk to internal occurs through collective electromagnetic processes rather than binary collisions. 
This is evident in a wide range of astrophysical systems, including supernova remnants (SNRs) \cite[e.g.,][]{morlino+12,caprioli12}, winds and jets of active galactic nuclei \cite[e.g.,][]{dermer+09,matthews+19, ajello+21}, heliospheric shocks  \cite[e.g.,][]{mason+99,turner+21}, and galaxy clusters \citep{brunetti+14}. 


The ability of a shock to accelerate ions to non-thermal energies is strongly dependent on the magnetic field inclination $\theta$ with respect to the shock normal. 
Specifically, at quasi-parallel shocks—where $\theta$ is small—the spontaneous generation of back-streaming energetic ions sets up a positive feedback loop, where these particles drive turbulence that, in turn, enhances further acceleration (\cite{caprioli+14a, caprioli+15}). 
These ions interact with the upstream flow, generating magnetic perturbations that promote diffusion across the shock and enable particles to gain energy through diffusive shock acceleration (DSA), resulting in characteristic power-law distributions \citep{bell78a, caprioli+20}.

On the other hand, oblique ($\theta > 45\deg$) and quasi-perpendicular shocks ($\theta \approx 90\deg$) are other particularly intriguing cases for particle acceleration. 
In the purely perpendicular case, ions can only penetrate the upstream region while gyrating around the ordered magnetic field, and
they cannot drive self-generated waves beyond one gyroradius from the shock.
This process of repeatedly crossing the shock surface while gyrating around the magnetic field, known as shock drift acceleration (SDA) \cite[e.g.][]{ball+01}, can significantly enhance particle energy. 
Although ion injection has been observed in kinetic simulations of non-relativistic shocks \citep{decker+85, jones+91, burgess+05, park+13, guo+14a, orusa+23}, a detailed analysis of this process at perpendicular shocks is still missing and is the main focus of this work.

In the past, fully kinetic particle-in-cell (PIC) simulations of (quasi-)perpendicular shocks did not show the presence of non-thermal ions. Previous works were realized in 1D \cite[e.g.,][]{leroy+82,shimada+00,scholer+03a,kumar+21,xu+20}, 2D \cite[e.g.,][]{amano+09a,lembege+09,kato+10,matsumoto+15,bohdan+21}, and 3D \cite[e.g.,][]{matsumoto+17}, though in relatively small boxes.
In our previous work (\cite{orusa+23}, hereafter Paper I) we presented self-consistent hybrid simulations (with kinetic ions and fluid electrons) that showed spontaneous (i.e., without pre-energized seeds and/or turbulence) ion acceleration at quasi-perpendicular shocks, a result obtained only in 3D simulations. 
These findings have been applied to experiments with astrophysically-relevant conditions in \cite{orusa+25a}.

Here we perform 2D and 3D hybrid perpendicular shock simulations with a specific focus on how particles are injected, and why 3D is necessary in order to achieve efficient injection and acceleration. The paper is structured as follows: in Section \ref{sec:simulation_setup}, we describe the simulation setup; in Section~\ref{sec:results}, we examine how our results depend on the adopted resolution and discuss the physics of ion injection in perpendicular shocks; finally, in Section~\ref{sec:conclusions}, we present our conclusions.

\section{Simulation setup}\label{sec:simulation_setup}

The simulations in this study were performed with {\tt dHybridR} \citep{haggerty+19a}, a highly parallel, hybrid simulation code, here used in the non-relativistic regime \citep{gargate+07}. The code treats ions kinetically while electrons are modeled as a charge-neutralizing fluid, following an adiabatic equation of state \citep[see][for an extended discussion on how to model it in shocks]{caprioli+18}. 

In contrast to PIC simulations, the hybrid approach does not resolve the electron plasma scales. Therefore, it is possible to simulate larger systems without losing important information about the shock dynamics, which is mostly controlled by the ions. 
The main disadvantage is that, in the absence of constraints imposed by electron-scale physics, it is not straightforward to determine the resolution required to accurately capture ion-scale dynamics.
We will discuss the appropriate resolution to adopt in Section \ref{sec:simulation_res}, and explain why, once the ion skin depth is properly resolved, smaller-scale structures become irrelevant for capturing ion-scale physics.

Length scales in the simulation are expressed in units of the ion skin depth, $d_i=c/\omega_p$, where $c$ is the speed of light and $\omega_p=\sqrt{4\pi n e^2/m}$ is the ion plasma frequency, with $m$, $e$, and $n$ representing the ion mass, charge, and upstream number density, respectively. Time is measured in units of inverse cyclotron frequency $\omega_c^{-1}=mc/eB_0$, with $B_0$ being the strength of the upstream magnetic field. Velocities are normalized to the Alfvén speed, $v_A = B_0/ \sqrt{4\pi m n} = c \,\omega_c/\omega_p$. The simulations capture all three components of the particle momentum and of the electric and magnetic fields.

The shock is generated by sending a supersonic flow toward a reflecting wall located at the left boundary \citep{caprioli+14a}. The interaction between the initial and reflected streams creates a shock that propagates to the right. In the downstream region the fluid is stationary, i.e., the simulation is performed in the downstream frame.
At the shock, the ordered bulk kinetic energy of the upstream flow is converted into thermal energy.
In our setup $\mathbf{v}_{sh} = - v_{sh} \mathbf{\hat{x}}$ is the upstream fluid velocity in the downstream reference frame. 
The initial magnetic field $\mathbf{B}_0$ makes an angle $\theta$ with the shock normal, which is oriented along the $\mathbf{\hat{x}}$-axis. We consider only purely perpendicular shocks with an angle $\theta=90^\circ$ and $B$-field directed along the $\mathbf{\hat{y}}$ axis. Ions are initialized with a thermal velocity $v_{th} = v_A$, corresponding to an initial temperature of $T_0 = \frac{1}{2} m v_A^2 /k_B$, where $k_B$ is the Boltzmann constant. Electrons are in thermal equilibrium with ions,  $T_e = T_i = T_0$. 
The sound speed is thus given by $c_s = \sqrt{2 \gamma k_B T_0/m}$, and the sonic Mach number is $M_s = M_A /\sqrt{\gamma}$, where $\gamma=5/3$ is the adiabatic index. 
This is equivalent to a plasma $\beta$ (ratio between thermal and magnetic pressure) of 2. 

\begin{table}[b]
\begin{center}
\caption{Left: Parameters for the runs in the Paper. Right: Corresponding acceleration efficiency $\varepsilon$ and energy spectral index $\alpha$ measured at $t=32 \omega_c^{-1}$.}
\label{tab::Fit_results_pp_pion}
\begin{tabular}{ c c c c | c c}
 \hline \hline
 Run & $M_{A}$ & $\Delta t$[$\omega_c^{-1}$] & $\Delta x$ [$d_i$] & $\varepsilon (> 10 E_{sh})$ & $\alpha$\\
\hline 
A   &   30   &   5 $\times 10^{-4}$ & 0.1 & 0.3\% & 6\\
B   &   60   &   2 $\times 10^{-4}$ & 0.1 & 0.6\% & 5\\
C   &   100  &   3 $\times 10^{-4}$ & 0.4 & 10\% & 2.5\\
D   &   100   &  1 $\times 10^{-4}$ & 0.1 & 0.9\% &  4\\
 \hline \hline
\end{tabular}
\end{center}
\end{table}

We define a characteristic energy scale, $E_{sh}$, as the bulk kinetic energy of the incoming upstream ions:
$E_{sh} = \frac{1}{2} m v_{sh}^2 = \frac{1}{2} m M_A^2 v_A^2$. 

The simulations use different numbers $N$ of cells per $d_i$, as discussed in Section \ref{sec:simulation_res}, which results in varying the spatial resolution $\Delta x$. The transverse box extends over 20 $d_i$ in the transverse directions $y$ and $z$, and we adopt 8 ion particles per cell. The timestep $\Delta t$ is chosen in each simulation to satisfy the  Courant-Friedrichs-Lewy (CFL) condition. We verified the robustness of our results through convergence tests in number of particles per cell, box size, and temporal resolution for each spatial resolution considered.
Details of the simulations are provided in Table \ref{tab::Fit_results_pp_pion}.
Most of our work is based on 3D simulations, but we also use 2D simulations for comparison. All 2D simulations are performed with $\mathbf{B}_0$ lying in the simulation plane, since we showed in Paper I that when it is out of plane, the field undergoes simple compression at the shock, as in a laminar magnetohydrodynamic (MHD) shock.

\section{Results}\label{sec:results}
\subsection{Simulation resolution}\label{sec:simulation_res}

\begin{figure}[t]
    \begin{center}\includegraphics[width=0.46\textwidth, clip=true,trim= 60 0 50 0]{"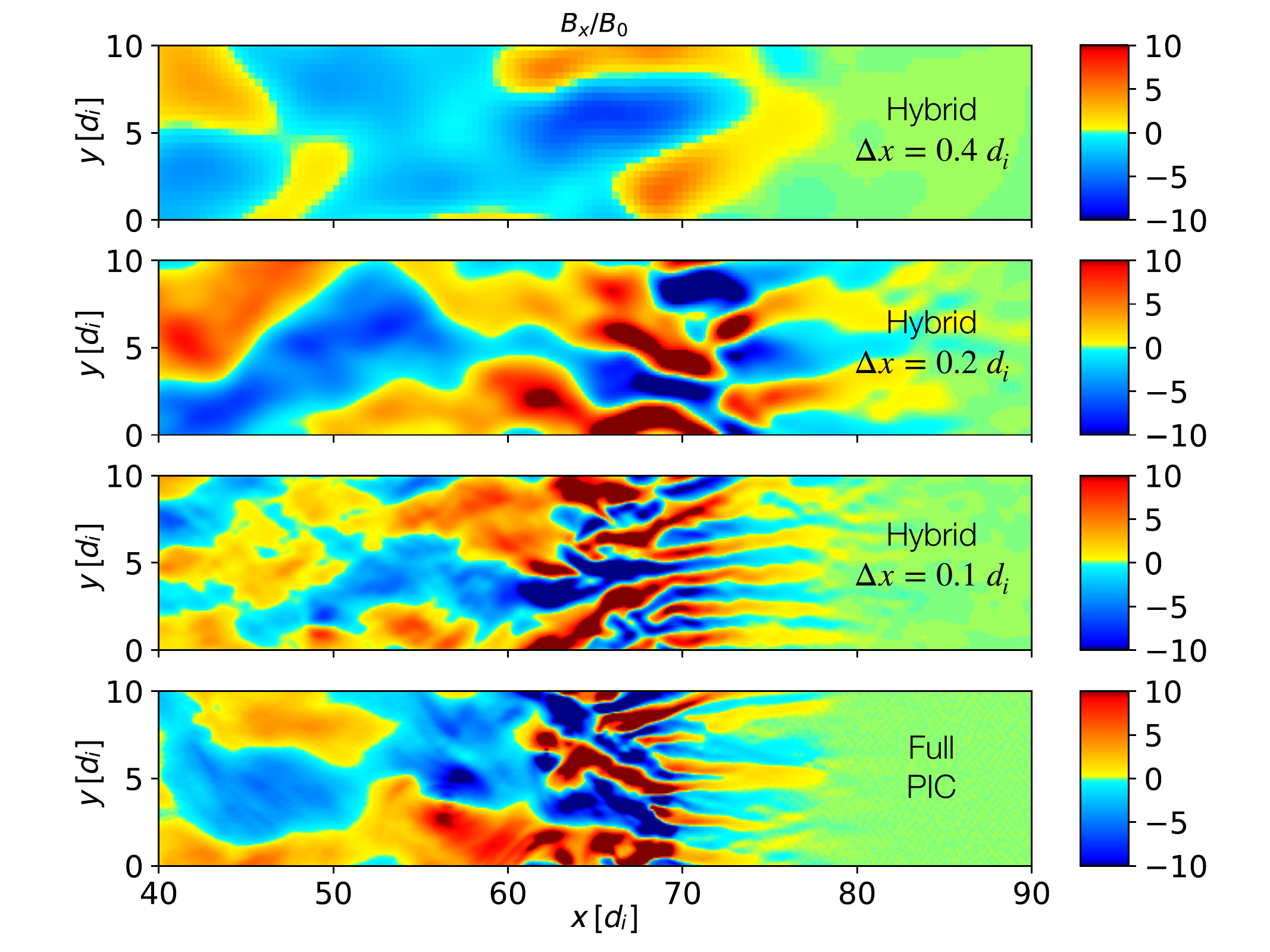"}
    \caption{$B_x$ component of the magnetic field from simulations of a shock with $M = 30$ at $\theta = 90^\circ$, obtained using a low resolution hybrid (first panel), a medium resolution hybrid (second panel), a high resolution hybrid (third panel) and a full PIC simulation (fourth panel) with $m_i/m_e = 49$, performed using Tristan-MP \citep{spitkovsky05}.} 
    \label{Fig:magnetic_PIC_hybrid}
    \end{center}
\end{figure}

We start our discussion with the analysis of the dependence of the non-thermal ion tail on the simulation resolution and the comparison of our hybrid results with full PIC simulations.

One common assumption is that resolving sub-$d_i$ turbulence may not be necessary, especially since the gyroradius of the thermal ions reflected at the shock and gyrating in the upstream magnetic field is on the order of $\sim M_A d_i$, and even larger for ions accelerated through SDA.
The spatial resolution is particular taxing from the computational point of view in 3D, since it constrains also the time resolution: eventually the computational cost scales as $N^4$.
The parameters of the simulations presented here are in Table \ref{tab::Fit_results_pp_pion}. We consider two resolutions,  $\Delta x=0.4$ and 0.1 $d_i$. The former corresponds to the one used in Paper I and the latter is the one necessary to achieve a visual agreement between a 2D hybrid simulation with $M_A=30$ and its fully kinetic counterpart performed with Tristan-MP \citep{spitkovsky05}, using a mass ratio $m_i/m_e =49$. We have also tested a higher mass ratio, $m_i/m_e = 256$, which yields results that are nearly identical at ion kinetic scales. Nevertheless, we note that even in full-PIC simulations some caution is warranted when interpreting the results, as residual dependencies on the adopted mass ratio may persist. 
This comparison, shown in Figure~\ref{Fig:magnetic_PIC_hybrid}, serves to validate resolution convergence by benchmarking the hybrid approach against simulations that include the full electron kinetic physics. 
Figure~\ref{Fig:magnetic_PIC_hybrid} shows also the result obtained from a 2D simulation at $M_A=30$ with an intermediate resolution of $\Delta x=0.2$ $d_i$. 
The resolution of $\Delta x=0.1$ $d_i$ accurately captures the filaments produced by the ion-Weibel instability \citep{weibel59,bohdan+21,jikei+24,orusa+26}  that mediates the shock transition. These filaments are subsequently compressed and amplified by the shock, a process not properly captured at lower or intermediate resolutions.

\begin{figure}[t]
    \includegraphics[width=0.46\textwidth]{"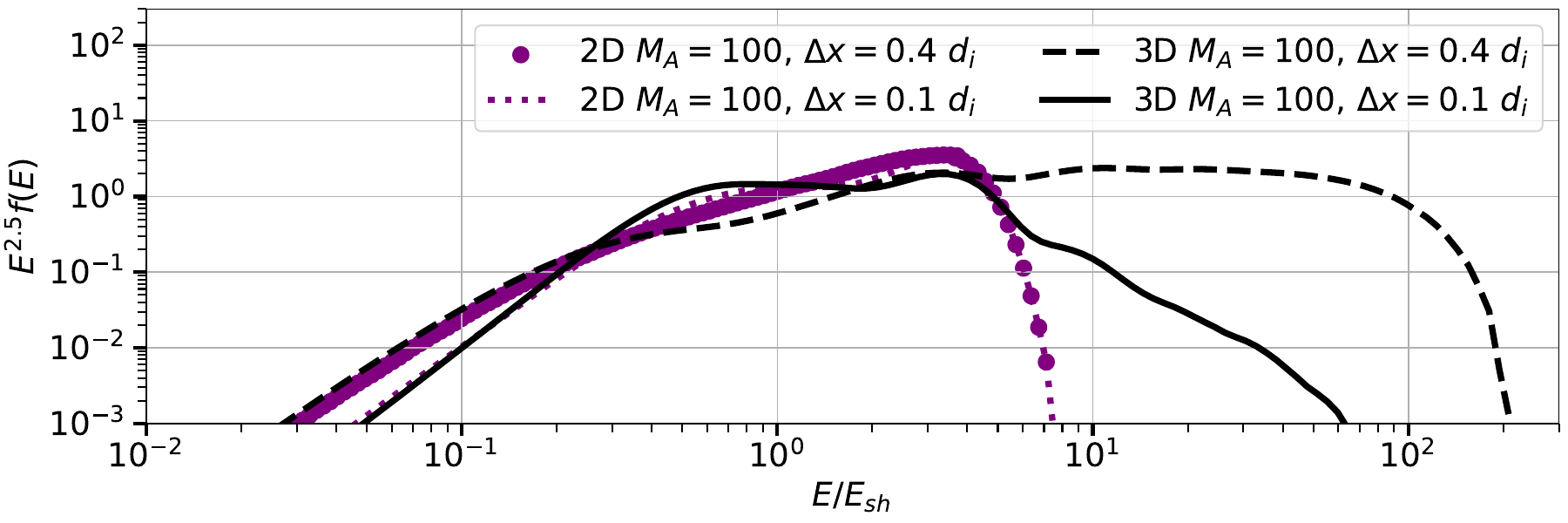"}
    \includegraphics[width=0.46\textwidth]{"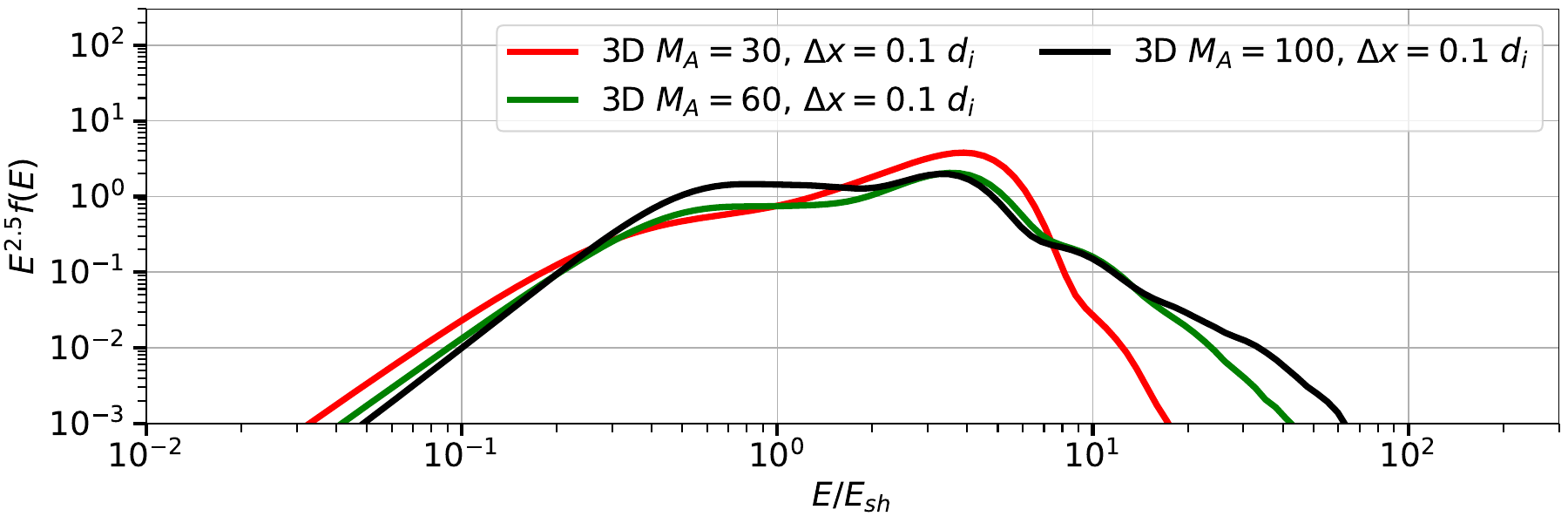"}
    \caption{Downstream ion energy spectra from various simulations. The top panel shows the spectra for $M_A = 100$ in both 2D and 3D, comparing two resolutions ($\Delta x = 0.4$ and $0.1 \, d_i$) at $t = 32 \, \omega_c^{-1}$. The bottom panel displays spectra for $M_A = 30$, 60, and 100, all with a resolution of $\Delta x = 0.1 \, d_i$  at $t = 32 \, \omega_c^{-1}$.
    } 
    \label{Fig:Spectrum_low_high_res}
\end{figure}

When comparing the high- and low-resolution downstream energy spectra in Figure~\ref{Fig:Spectrum_low_high_res} 
for 2D and 3D simulations with $M_A = 30, 60$ and $100$, we point out two key aspects. 
Up to  $E \sim 6-7 \, E_{sh}$ in both 3D and 2D (as shown in the upper panel of Figure~\ref{Fig:Spectrum_low_high_res}), the spectra appear nearly identical, featuring a distinct bump around $5 E_{sh}$ for all cases. 
This similarity suggests that the efficiency of thermal particle reflection at the shock is comparable in both 2D and 3D, across different resolutions. Once reflected, the particles gyrate in the upstream motional electric field directed along positive $\mathbf{\hat{z}}$, gaining energy during their first gyration, corresponding to the initial cycle of SDA. 

Ions with energies higher than the bump at $5 E_{sh}$ are consistently present only in 3D. However, in the higher-resolution cases, the resulting spectra are steeper. 
The number of injected particles is then not only a function of the dimensionality of the spatial domain (i.e., 2D vs 3D), but also depends on the characteristics of the small-scale turbulence in the post-shock region, which controls the probability of return of particles from downstream, as we will discuss in the following \citep{bell78a, bell+11}. 
Therefore, the results presented in Paper I remain  qualitatively valid:
1) 3D is essential to prevent advection after just one SDA cycle, as shown in the upper panel of Figure~\ref{Fig:Spectrum_low_high_res}, while all the 2D setups fail to accelerate particles, regardless of the resolution; 
2) ion acceleration depends on $M_A$ and higher-$M_A$ shocks exhibit systematically flatter spectra, as shown in the lower panel of Figure~\ref{Fig:Spectrum_low_high_res}. However, determining the exact dependence of the SDA slope and the acceleration efficiencies on $M_A$ requires higher resolution than the one we used in Paper I.

The choice of numerical resolution is a critical aspect of hybrid-PIC simulations. One may wonder whether further increasing the resolution would suppress particle acceleration or lead to more reliable results. This is not necessarily the case. The resolution adopted here ($\Delta x = 0.1\, d_i$) was chosen because it reproduces the ion-scale magnetic-field structure observed in corresponding full-PIC simulations, which we use as a benchmark for physical fidelity. Increasing the resolution beyond this range does not automatically improve the accuracy of the hybrid approach. Instead, it can introduce numerical artifacts, most notably the onset of non-physical whistler modes due the fact that hybrid assumes massless $e^-$, which prevent convergence \citep{orusa+26}. This behavior limits the range of resolutions over which hybrid simulations provide a faithful description of the ion-scale instabilities relevant to quasi-perpendicular shocks.


\begin{figure}[t]
\begin{center}
    \includegraphics[width=0.46\textwidth]{"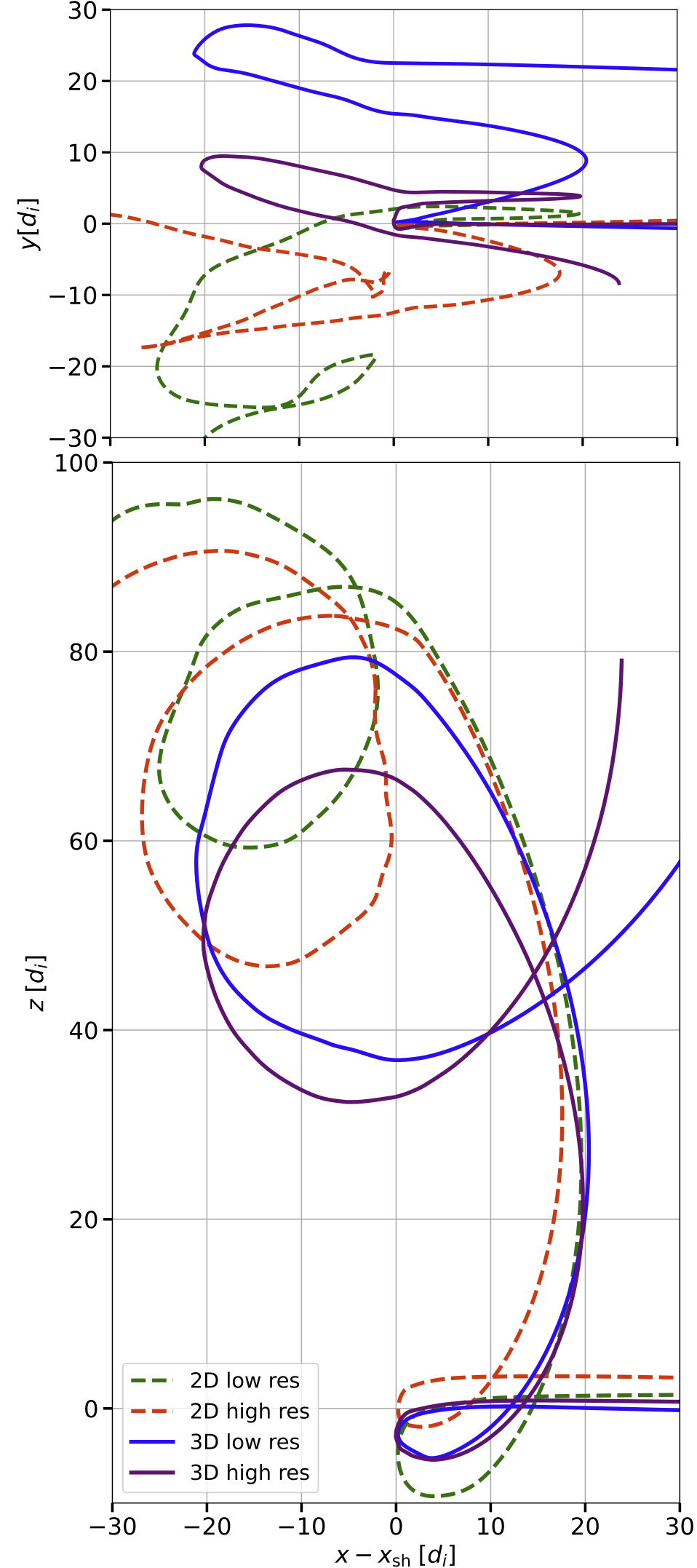"}
    \caption{Trajectories in the $x-y$ plane (top) and $x-z$ plane (bottom) of one representative particle per simulation—both 2D (dashed) and 3D (solid), at high and low resolution of $M_A=30$ shock—that exhibit similar behavior in the initial stages of their evolution (see the text for details).}
    \label{Fig:doppleganger}
    \end{center}
\end{figure}

\subsection{The life of an accelerated particle in a perpendicular shock}\label{sec:particles}

\begin{figure}[t]
\begin{center}
    \includegraphics[width=0.49\textwidth]{"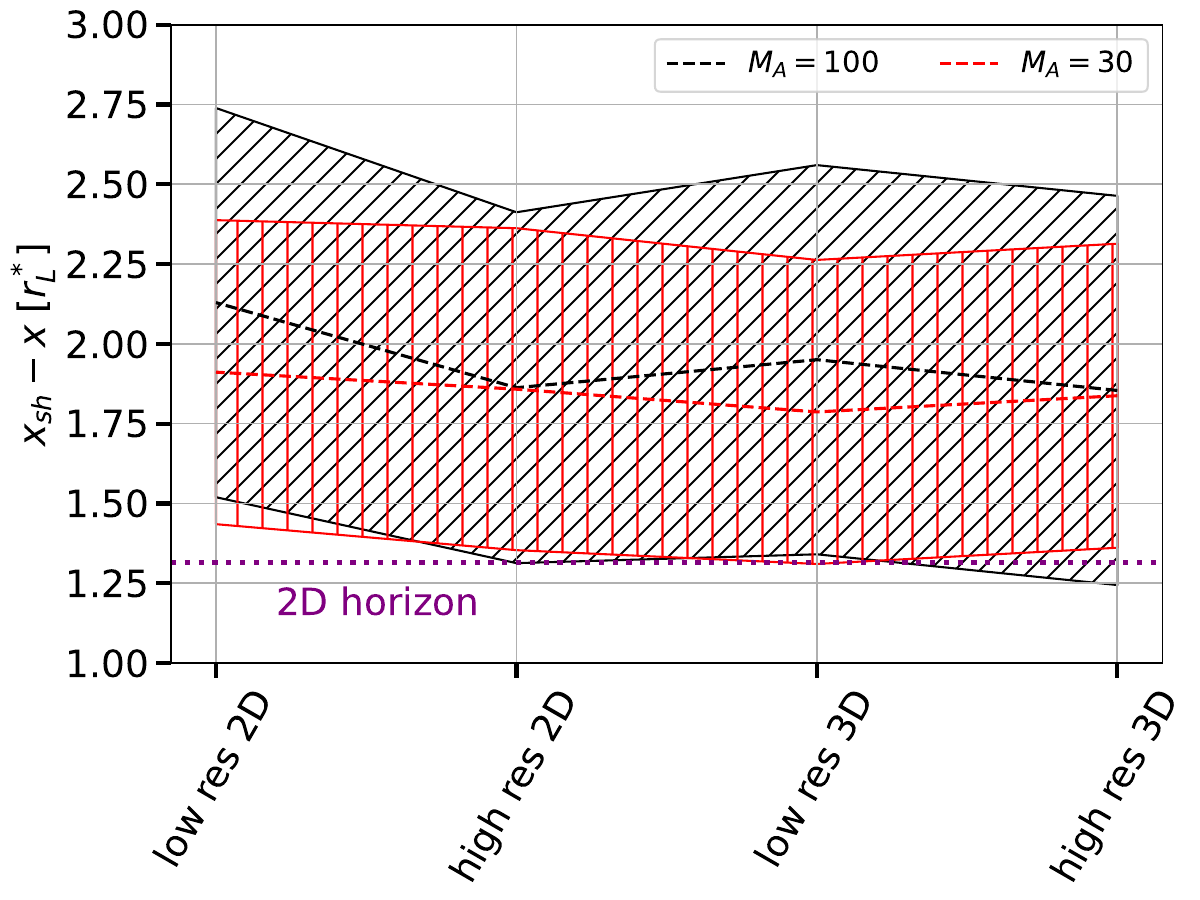"}
    \caption{The average penetration depths of ions with energy $E = 7.5 E_{sh}$ are shown for $M_A = 30$ (red lines) and 100 (black lines) in units of $r_L^*=r_L(7.5 E_{sh})$ (see the text for details). These represent the distances traveled downstream where $p_x$ changes sign. They are reported for 2D and 3D simulations at low and high resolutions. The hatched bands mark the $2\sigma$ intervals of the penetration distance distributions. The dotted line represents the 2D return horizon, as defined in Section \ref{sec:advection_horizon}.
    } 
    \label{Fig:penetration}
    \end{center}
\end{figure}

To investigate the differences in particle injection between 2D and 3D simulations at both low and high resolutions, we present in Figure~\ref{Fig:doppleganger} the $x$–$y$ (top) and $x–z$ (bottom) trajectories of one representative particle from each simulation,comparing 2D and 3D, both at high and low resolution for $M_A = 30$. These particles are selected because they exhibit similar early-time behavior, including comparable distances traveled ahead of the shock after the first reflection, similar energies after the first gyration, low $y$-momentum following the initial cycle, and similar maximum penetration depths into the downstream region. These “doppelgänger” particles serve as analogs across different simulation setups, enabling a direct comparison of their dynamics.

\begin{figure*}[t]
\begin{center}
    \includegraphics[width=0.49\textwidth]{"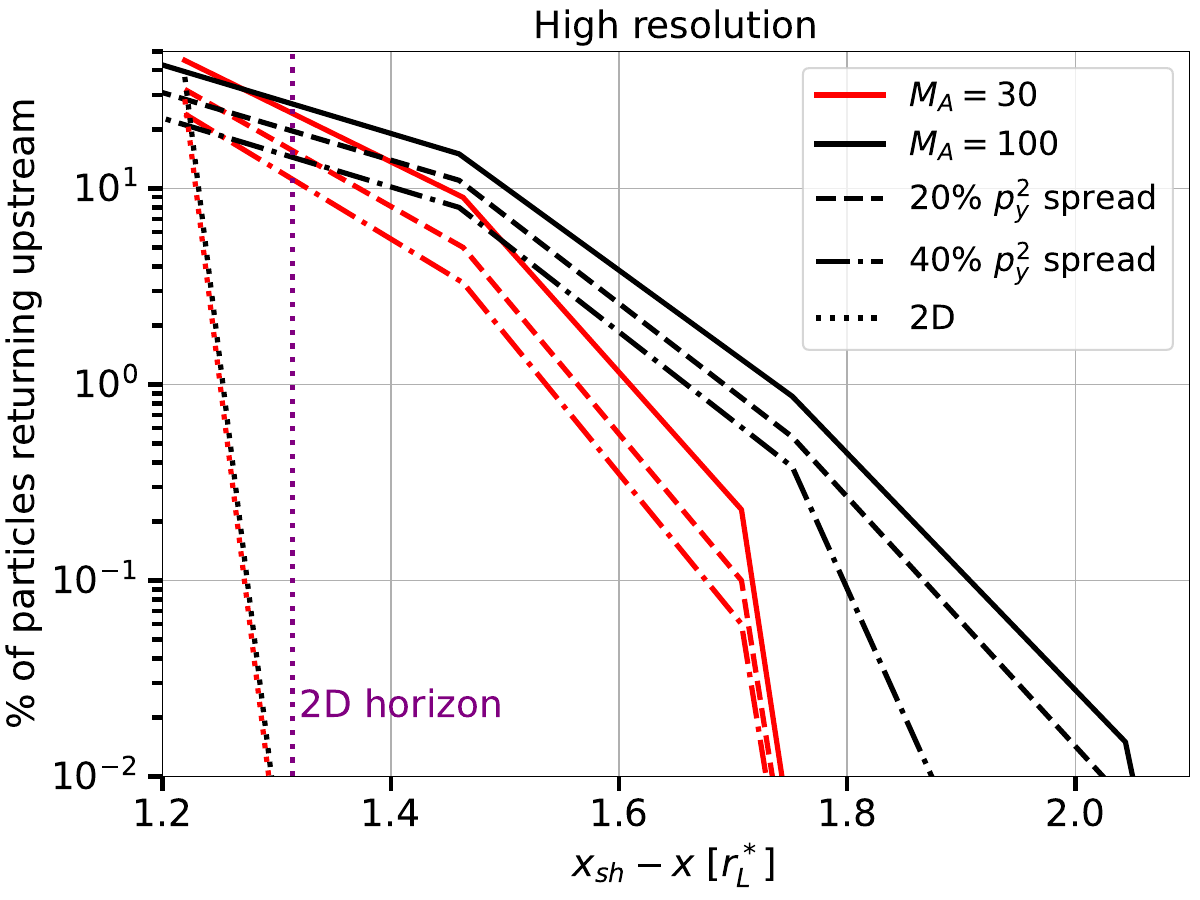"}
    \includegraphics[width=0.49\textwidth]{"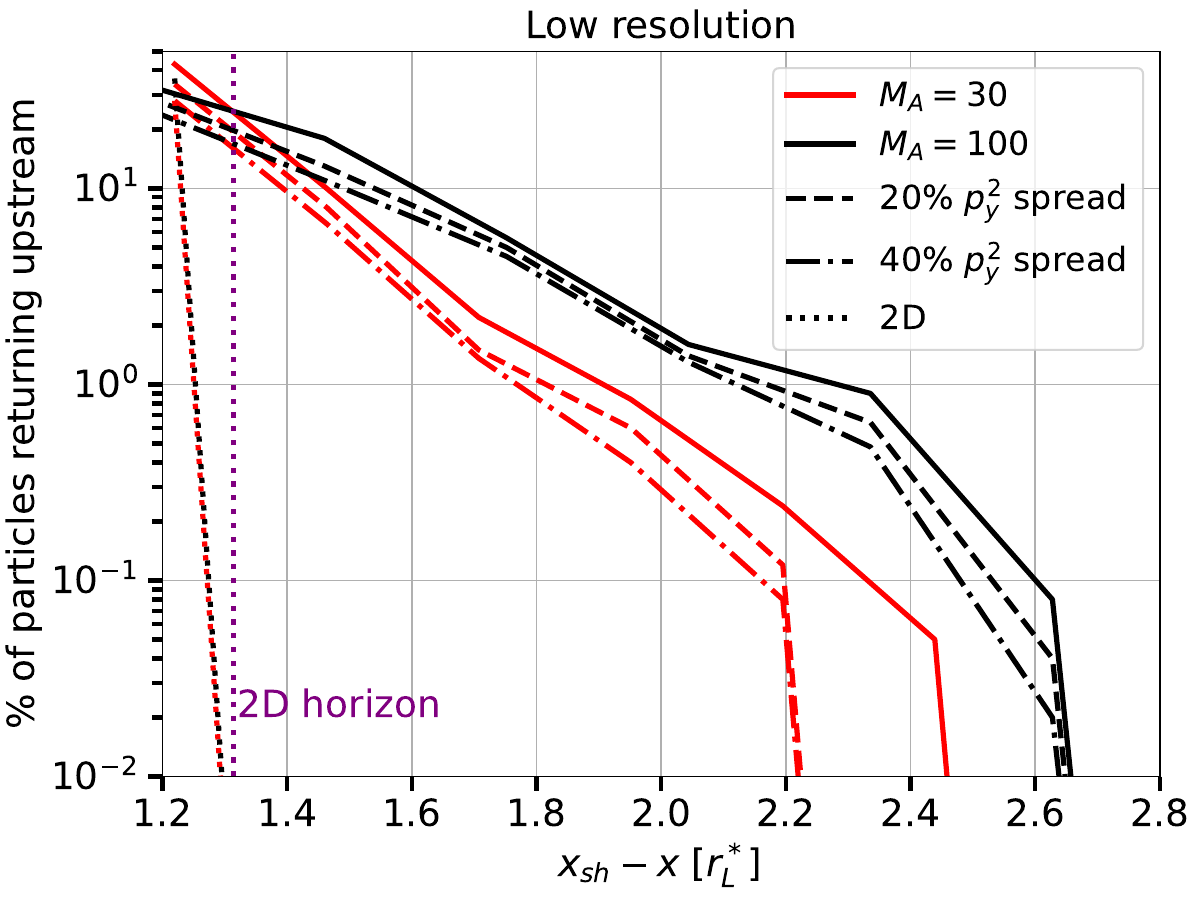"}
    \caption{Percentage of particles with energy $E=7.5 E_{sh}$ that, starting from various initial positions behind the shock, expressed in units of $r_L^*=r_L(7.5 \, E_{sh})$, are able to return to the upstream and reach a distance in front of the shock of $r_L^*/4$. This is presented for 2D and 3D simulations at both high (left panel) and low (right panel) resolutions. The initial particle momentum is either along $z$ or includes some momentum along the $y$-direction drawn from a uniform distribution with limits $p_y^2/p^2=[0.0, 0.2]$ or $p_y^2/p^2=[0.0, 0.4]$.
    } 
    \label{Fig:horizon}
    \end{center}
\end{figure*}

Each particle undergoes a first reflection at the shock and propagates a similar distance upstream, between $17$ and $20$ $d_i$, gaining energy from the motional electric field. Following their initial reflection, they predominantly gyrate in the $x–z$ plane due to the background magnetic field $\mathbf{B_0}=B_0 \mathbf{\hat{y}}$, with a positive momentum $p_z$ and a small $p_y$, reaching energies around $7 E_{sh}$ in this first SDA cycle. 
After this stage their evolutions begin to diverge. Once they penetrate in the downstream, all particles travel comparable distances before their $p_x$ switches from negative back to positive;
we call this distance, which ranges between 20 and 25 $d_i$ downstream of the shock, the maximum penetration depth.

The particles then attempt to return upstream, but in 2D they are prevented to do so by strong magnetic field fluctuations that act as an effective barrier. 
In contrast, in 3D, particles are able to re-enter the upstream region, allowing the SDA acceleration process to proceed for a second cycle. We define these particles as being injected into the acceleration process. As seen from their trajectories, particles in 3D experience minimal scattering after their initial reflection and gyration: their motion largely follows regular gyration patterns without significant deviations, indicating an absence of strong scattering events during their return to the upstream region.

In what follows, we present quantitative evidence that the magnetic turbulence behind the shock exhibits different \textit{porosity} in 2D and 3D. Here, porosity refers to the ability of the post-shock region to allow particles to traverse it and return upstream without being blocked by the magnetic barrier. This difference, which manifests during the first downstream gyration, is the key distinction between 2D and 3D configurations.

\subsection{The return horizon}\label{sec:advection_horizon}

To assess this porosity argument, we perform two experiments in which test particles (40000 particles in each test) are propagated through the electromagnetic fields extracted from the simulations at different times (8, 16, 24, and 32 $\omega_c^{-1}$) which yield consistent results with one another. The motion of the shock is accounted for by translating the shock position at each timestep, so that particles experience the correct relative dynamics. We use ions with energy $E = 7.5\,E_{sh}$; 
particles with this energy have already undergone the first SDA cycle and above this  threshold differences between 2D and 3D spectra become significant. 

In the first test, we estimate the distance these particles travel into the downstream region before their $p_x$ changes sign, which we define as their \textit{penetration depth}. In the second test, we determine the critical distance from the shock beyond which the fraction of particles returning upstream drops to zero, indicating that return/injection becomes effectively impossible beyond this point. We refer to this distance as the \textit{return horizon}.
If in 2D the average penetration depth exceeds this return horizon, while in 3D it does not, this effect can be identified as the key distinction between the two geometries.

We begin with the first test. Due to the upstream fields, reflected thermal particles predominantly gyrate in the $x – z$ plane and enter the shock with large $p_{x,z}$ components and a smaller $p_y$ after the first SDA cycle. 
We initialize test particles at the shock and model their momentum by sampling $p_x^2$, $p_y^2$, and $p_z^2$ from uniform distributions. The sampling intervals are based on the phase spaces measured in the simulations; all test particles are constrained to have the same energy of $E = 7.5\,E_{sh}$. 
Specifically, we use $p_y^2/p^2 \in [0.0\,; 0.2]$ with both positive and negative $p_y$, $p_x^2/p^2 \in [0.0\,; 0.4]$ with $p_x < 0$ (since the particle is entering the shock), and $p_z > 0$ obtained as $p_z=\sqrt{p^2-p_x^2-p_y^2}$.

We introduce a reference Larmor radius, $r_L$, that depends on $E$ and $M_A$,
defined as the Larmor radius of a particle with energy $E$ in a magnetic field equal to four times the background field $B_0$. This value corresponds to the downstream magnetic field expected from a laminar MHD shock with a density compression ratio of four. The expression for $r_L$ is:
$r_L = \frac{M_A}{4} \sqrt{\frac{E}{E_{sh}}} \; d_i$.
\\
This analysis, reported in Figure~\ref{Fig:penetration}, shows that particles reach their downstream penetration depth (expressed in units of $r_L^*=r_L(7.5 \, E_{sh})$) 
between $1.25 \; r_L^*$ and $2.25 \; r_L^*$ for $M_A = 30$, and between $1.25 \; r_L^*$ and $2.75 \; r_L^*$ for $M_A = 100$. Red and black lines indicate the average penetration distances for $M_A = 30$ and $M_A = 100$ respectively, while the hatched bands represent the $2\sigma$ intervals of the penetration distributions. These values exhibit only minor differences between 2D and 3D simulations and between the two spatial resolutions we explored.


After reaching their maximum penetration depth, particles continue their motion and attempt to return upstream. To investigate this process and determine the return horizon, we perform test-particle simulations with $E = 7.5\,E_{sh}$ to quantify the probability of returning upstream as a function of the distance behind the shock from where they start. 

To identify this horizon, we consider particles initialized at their maximum penetration depth, where $p_x = 0$, and assess how the return probability depends on the starting distance behind the shock. We explore two scenarios: one in which the particle momentum is aligned along the $z$-direction, and another in which particles have a finite spread in $p_y$.
In the latter case, $p_y^2$ is sampled from a uniform distribution and normalized such that the total particle energy is set at $E = 7.5 , E_{sh}$, with $p_y^2/p^2 = [0.0 \; 0.2]$ or $p_y^2/p^2 = [0.0 \; 0.4]$ used as limiting values. We introduce a finite spread in the $y$-direction to demonstrate that, regardless of the initial momentum distribution, the return horizon remains fundamentally different between 2D and 3D.

We then count how many of these particles are able to cross the shock, re-enter the upstream region, and reach a distance of at least $r_L^*/4$ ahead from the shock front. This distance is arbitrarily chosen to ensure that only particles actually propagating into the upstream are counted, excluding those whose trajectory barely reaches the shock.

Figure \ref{Fig:horizon} shows the percentage of particles with initial energy $E = 7.5 \, E_{sh}$ that, starting from various distances behind the shock, are able to return upstream. Results are presented for both 2D and 3D simulations at high (left panel) and low (right panel) resolution for $M_A = 30$ and $M_A = 100$. 
A clear contrast emerges: in 2D, the return probabilities rapidly drop to zero at the return horizon situated at $1.31 \, r_L^*$ for both values of $M_A$, indicating a strong advective effect that prevents particles from escaping the downstream region. In 3D, however, a significant fraction of particles starting  well beyond $1.31 \, r_L^*$ are able to return upstream, consistent with the behavior observed in the ``doppelgänger'' particle trajectories of Section~\ref{sec:particles}.
The 2D return horizon approximately coincides with the lower boundary of the $2\sigma$ band of the distribution of penetration depths, implying that most particles propagate downstream to distances beyond the return horizon, as shown in Figure~\ref{Fig:penetration}. 
This explains the inability of particles to return upstream in 2D.
\begin{figure*}[t]
\begin{center}
    \includegraphics[width=0.89\textwidth]{"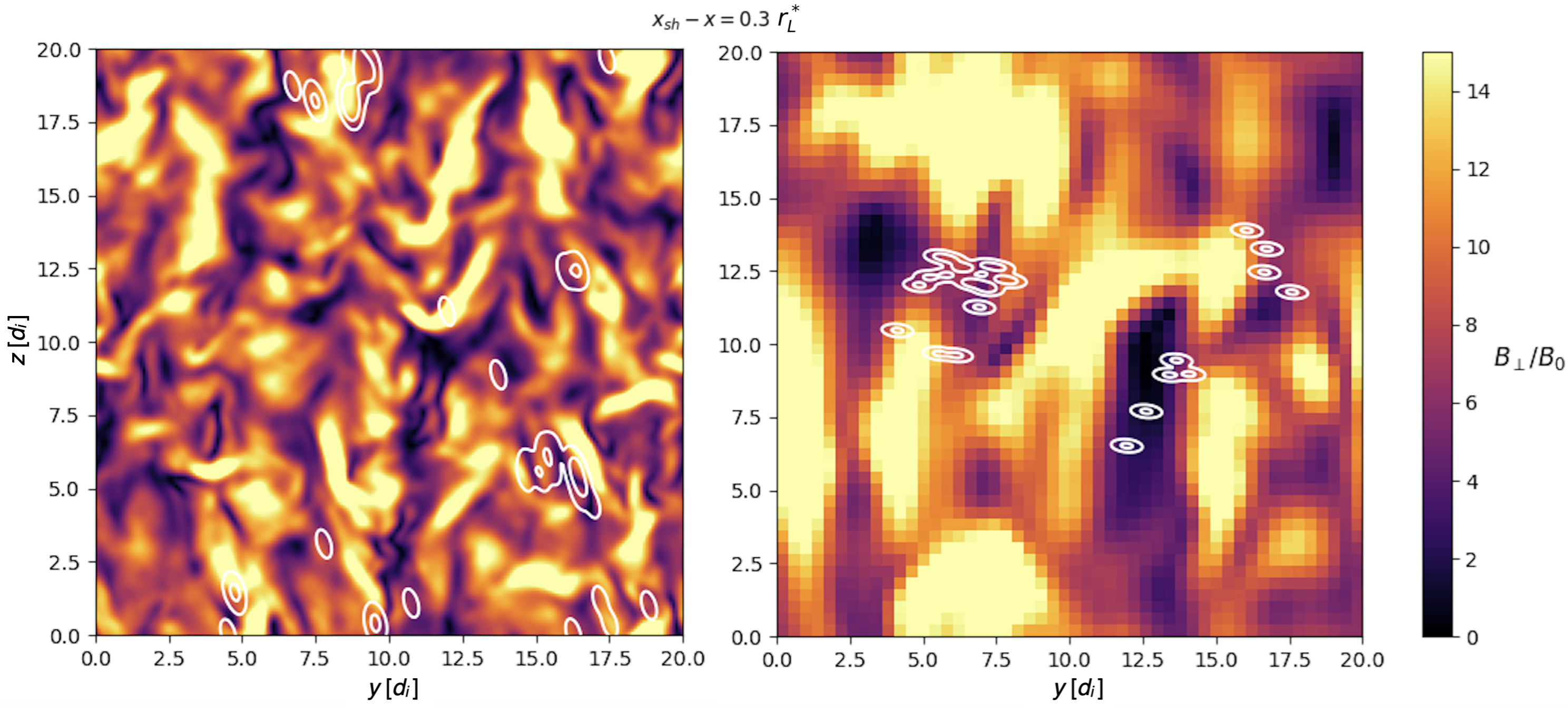"}
    \caption{Spatial distribution (white contours) in the $y$–$z$ plane of test particles from our horizon exercise (based on the setup used for Figure \ref{Fig:horizon}) that successfully return upstream. The snapshot is taken at a position 0.3 $r_L^*$ behind the shock (with the test particles initially located at 1.6 $r_L^*$), for both high-resolution (left) and low-resolution (right) simulations at $M_A=30$. This distribution is compared to a snapshot of $B_{\perp}$ at the same location, illustrating that particles capable of returning upstream tend to avoid regions of strong $B_{\perp}$.
    } 
    \label{Fig:particle_position}
    \end{center}
\end{figure*}

The distributions in Figure~\ref{Fig:horizon} extend to greater distances in low-resolution simulations compared to high-resolution ones, i.e., particles can more easily propagate back upstream in lower-resolution runs. This explains the higher acceleration efficiency observed at lower resolution;
here efficiency is defined as the fraction of post-shock energy density carried by ions with energies above $10\,E_{sh}$ and it is reported in Table \ref{tab::Fit_results_pp_pion}. 
Likewise, increasing $M_A$ results in a larger return horizon when expressed in units of $r_L^*$, again consistent with the higher efficiency. 
The fact that spectra in Figure~\ref{Fig:Spectrum_low_high_res} are harder for lower resolution and higher Mach number can also be interpreted as a direct consequence of increased return probability \citep{bell78a, jones+98}; as more particles are injected into the acceleration process, the energy spectrum extends to higher energies and becomes flatter.
Introducing a finite spread in $p_y$, while keeping the total particle energy fixed at $E=7.5 E_{sh}$, reduces the return probability because it decreases the component of the particle’s momentum perpendicular to the mean magnetic field (which lies along the $y$-axis). A lower $p_\perp=\sqrt{p^2-p_y^2}$ implies a smaller Larmor radius, making it more difficult for particles to return upstream.

\subsection{The shock porosity}\label{Results}

To understand the origin of the different porosity in 2D versus 3D, we examine the downstream magnetic field structure. 
Since a field that is perpendicular to the shock normal can hinder particle propagation along $x$ (as required to go back upstream), we focus on $B_\perp = \sqrt{B_y^2 + B_z^2}$.
Thus, it is essential evaluate whether regions exist in the downstream where $B_\perp$ becomes sufficiently weak to allow particles to escape back upstream. A particle returning from the downstream and encountering a locally small $B_\perp$ effectively has a Larmor radius large enough to reach the upstream. 

\begin{figure*}[t]
\begin{center}
    \includegraphics[width=0.89\textwidth]{"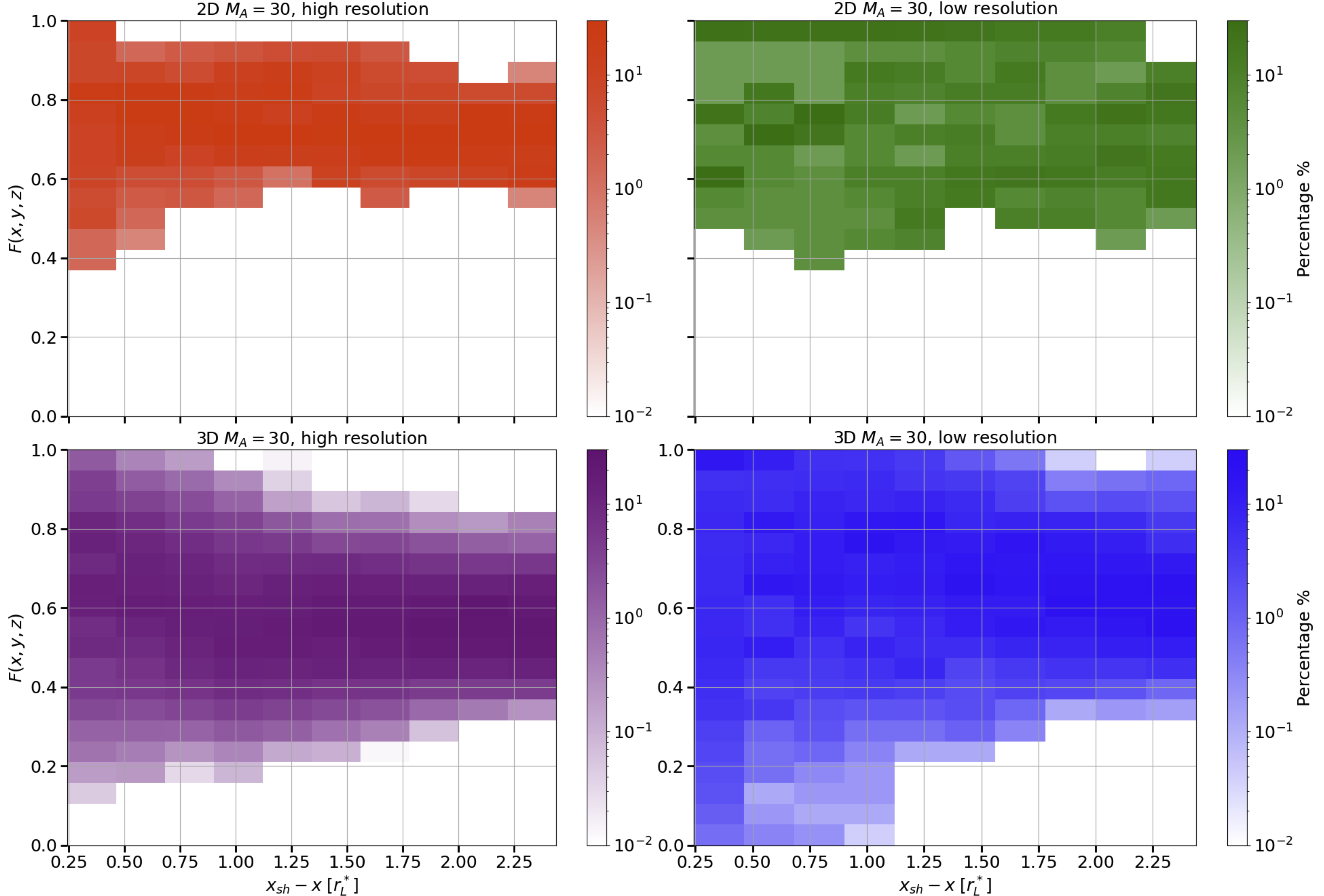"}
    \caption{Filling factor $F(x,y,z)$ distribution obtained for 2D (top row) and 3D (bottom row) at $M_A=30$ high (left) and low resolution (right) as a function of the downstream distance from the shock. $F(x,y,z)$ is computed at each $y$–$z$ coordinate for different $x$ from the shock. From this sample, the whole distribution is obtained. A small value of $F$ implies the presence of regions of weak $B_{\perp}/B_0$. The comparison clearly shows that in 3D, $F$ tends to be systematically lower than in 2D, indicating enhanced porosity.
    } 
    \label{Fig:fillingfactor}
    \end{center}
\end{figure*}




To support this argument, we consider test particles that successfully return upstream  from an initial distance of 1.6 $r_L^*$;
their spatial distribution in the $y$–$z$ plane at $x_{sh}-x= 0.3 \, r_L^*$ is shown in Figure~\ref{Fig:particle_position} as white contours, together with the slice of $B_{\perp}$ taken at the same location, for both high- and low-resolution simulations with $M_A = 30$. It is evident, especially for the low resolution simulation, that particles capable of returning upstream tend to avoid regions of strong $B_{\perp}$, instead favoring areas where $B_{\perp}$ is small.

To quantify the occurrence of such regions, we perform a statistical analysis of the filling factor of small-$B_{\perp}$ regions in the near downstream.
In particular, we assess the shock porosity by searching for extended regions—“channels” or “tubes”—at fixed $y$ and $z$ coordinates where $B_{\perp}$ remains consistently below a defined threshold along the $x$ direction, all the way to the shock. The presence of such structures implies reduced magnetic confinement and enables particle propagation back towards the shock. Although particle trajectories are fully three-dimensional and not confined to fixed $y$–$z$ locations, this test provides both qualitative and quantitative evidence for the existence of preferential escape pathways.

We define the following function:
\begin{equation}
    F(x,y,z) = 1 - \frac{\int_x^{x_{sh}} f(x',y,z)\, dx'}{x_{sh}-x},
\end{equation}

where the integrand function \( f(x',y,z) \) is defined as
\begin{equation}
f(x',y,z) = 
\begin{cases}
1 & \text{if } B_{\perp}(x',y,z)/B_0 < T, \\
0 & \text{otherwise}.
\end{cases}
\end{equation}
The threshold $T$ is chosen to be $T = \rho/\rho_0$, where the density $\rho$ is averaged over the $y$ and $z$ directions. This choice of $T$ corresponds to a shock-perpendicular magnetic field that merely results from flux freezing (i.e., excluding self-generated fluctuations) because of the density compression: $B_{\perp}/B_0 \sim \rho/\rho_0$. This definition of $F$ assesses the fraction of the line segment between position $x$ and the shock surface at $x_{sh}$ where the perpendicular magnetic field exceeds the chosen threshold. A small value of $F$ implies that a large portion of the path lies in regions where $B_{\perp}/B_0$ is weak, and consequently, the porosity is high.
 
Figure~\ref{Fig:fillingfactor} shows the distributions of $F$ values obtained for all ($y, z$) combinations, plotted as a function of $x$ for both 2D (top row) and 3D (bottom row) configurations, at high (left) and low (right) resolutions. 
The comparison clearly shows that in 3D, $F$ tends to be systematically lower than in 2D, indicating enhanced porosity.


In 2D, the translational invariance in the $z$ direction implies that the magnetic field is organized in "walls", with the magnetic structures forming extended, sheet-like barriers that span the simulation domain and systematically impede particle transport. In 3D, by contrast, the amplified magnetic field occupies a smaller volume fraction, leaving localized regions of weak field in which particles experience reduced magnetic deflection. In the low-resolution case, there are extended regions of very weak magnetic field, for example between $z \simeq 7.5$ and $15\,d_i$ in Fig.~\ref{Fig:alla_fields}. In the high-resolution simulation, while localized regions reach field strengths of order $\sim 15\,B_0$, there are also intervals (e.g., between $z \simeq 7$ and $14\,d_i$ in Fig.~\ref{Fig:alla_fields}) where the field is significantly weaker, including regions without strong amplification.
As a result, particles in 3D can propagate over distances comparable to their Larmor radius without encountering strong magnetic barriers. In contrast, in 2D a magnetic barrier inevitably intersects particle trajectories at some point, limiting their motion. This behavior is consistent with the trajectories shown in Fig. \ref{Fig:doppleganger} that are not affected by visible distortions of the orbits \citep{jones+98}.


Figure~\ref{Fig:fillingfactor} also justifies the spectral difference between low and high resolution simulations (see Figure~\ref{Fig:Spectrum_low_high_res}): in 3D, the energy spectrum becomes
significantly harder for low-resolution simulations. This difference is closely tied to the structure of the magnetic field: as shown in Figure~\ref{Fig:fillingfactor}, the low-resolution 3D case exhibits a statistically lower value of $F$ compared to its high-resolution counterpart. This finding is consistent with the larger return horizon (see Figure~\ref{Fig:horizon}) and higher fraction of injected particles observed in the low-resolution simulation.

This difference stems from the topology of the magnetic field downstream of the shock. Low-resolution simulations produce artificially large contiguous regions of reduced $B_{\perp}$ through which particles can easily propagate. In contrast, high-resolution simulations reveal a fine-scale, filamentary magnetic field structure produced by the ion-Weibel instability \citep{weibel59,bohdan+21,jikei+24}, with filaments approximately the size of the ion inertial length $d_i$ (see Figures \ref{Fig:magnetic_PIC_hybrid} and \ref{Fig:alla_fields}).
Along a given line of sight, these filaments form alternating regions of strong and weak $B_{\perp}$, as illustrated in Figure~\ref{Fig:alla_fields}. Since the filament size is $\sim d_i$, which corresponds to only $0.05\;r_L^*$, a given line at fixed y and z coordinates will likely encounter regions of strong fields. This diminishes the porosity benefits offered by the artificially-extended low-field regions seen in low-resolution simulations.

\subsection{The $M_A$ dependence}\label{sec:ma_dependence}

Another notable trend, independent of the simulation resolution, is the spectral hardening associated with increasing $M_A$. By analyzing the statistics of $B_{\perp}$, we find that the number of magnetic tubes satisfying the low-field condition (and so, allowing for efficient escape) decreases when increasing $M_A$, at a given $r_L^*$. This behavior is expected, since magnetic field amplification with respect to $B_0$ scales approximately as $\propto \sqrt{M_A}$ \citep{kato+10,bohdan+21,matsumoto+15}, implying that for a fixed threshold, the fraction of low-field regions diminishes.

While this might appear counterintuitive, suggesting fewer escape channels and thus a softer spectrum for higher $M_A$, one must account for an important competing effect. The stopping power of regions with strong $B_{\perp}$ also decreases with increasing $M_A$, since particles become less magnetized. Indeed, the Larmor radius in the background $B_0$ scales linearly with $M_A$, while the amplitude of magnetic field fluctuations grows more slowly, roughly as $\sqrt{M_A}$. It follows that the Larmor radius of particles gyrating in the amplified field scales as $\sqrt{M_A}$. As a result, strong field regions become less effective at halting particle injection, and the likelihood of particles returning upstream increases with $M_A$, ultimately leading to more efficient acceleration and harder spectra.

As demonstrated in Paper~I and in ~\cite{orusa+25a}, shocks with $M_A \lesssim 10$ do not efficiently accelerate particles, as evidenced by the lack of a non-thermal tail in the particle energy distribution. These low-$M_A$ shocks tend to settle into a laminar configuration, where the downstream magnetic field is simply compressed without developing appreciable fluctuations. In such configurations, particles are readily advected downstream and lack the conditions necessary to return upstream, effectively preventing acceleration in quasi-perpendicular shocks with high magnetization.


\section{Conclusions}\label{sec:conclusions}

In this work, we conducted an extensive study of the physics of ion injection and acceleration at perpendicular shocks using 2D and 3D hybrid simulations with different resolution and Mach number, analyzing the microphysics of ion injection into SDA. Our investigation reveals that ion acceleration can only happen in 3D, with the efficiency of particle injection closely linked to the intrinsic porosity of the 3D downstream magnetic turbulence. 
While the downstream region in 2D simulations appears to be composed of ``walls'' of strong $B_{\perp}$, a result of translational invariance in the $z$-direction, such structures are absent in 3D simulations. Instead, the 3D porosity is caused by the presence of low-$B_{\perp}$ regions in the downstream of the shock, where magnetic field lines can bend and twist in all directions, that are access points for particles to escape from the downstream and re-enter the upstream and continue the acceleration cycle.

We also observe that the shape of the particle energy spectrum in 3D depends on the simulation resolution. High-resolution simulations properly resolve short-wavelength magnetic fluctuations, capturing the fine-scale structures that particles interact with. In contrast, low-resolution simulations artificially enhance the shock's porosity, with more prominent regions of low $B_{\perp}$, leading to higher acceleration efficiency and artificially a harder spectrum (see Paper I). We determine the appropriate resolution by comparing the magnetic fields obtained from hybrid simulations with those from full PIC simulations. Regardless of resolution, however, particle acceleration remains an on–off process governed by the dimensionality of the simulation, with 2D simulations showing no sign of acceleration. 

By varying the shock Mach number, we show that higher $M_A$ shocks produce harder spectra and more efficient particle acceleration. Notably, our high-resolution simulations suggest that the threshold Mach number required to achieve a canonical $p^{-4}$ spectrum is likely underestimated in low-resolution studies (see Paper I). This reinforces the importance of both high spatial resolution and large simulation domains for properly capturing the nonlinear dynamics of perpendicular shocks.
We chose an angle of 90° in order to study the extreme, purely perpendicular limit, where the background magnetic field has no component along the shock normal. This configuration allows us to obtain a clean and controlled result, isolating the effects of perpendicular shocks without interference from parallel-field dynamics, and providing a clear benchmark for understanding particle acceleration and field amplification in this limit.
For oblique shocks, which are not covered in this work, the initial acceleration mechanism would still involve SDA. However, in this case, particles could escape upstream by streaming along the magnetic field lines, which is not possible in purely perpendicular shocks. The resulting escaping current could amplify the magnetic field, triggering a second phase of evolution where particles transition from SDA to DSA. This scenario will be explored in a future paper.

\begin{acknowledgments}
L.O. thanks Siddhartha Gupta and Robert Ewart for insightful discussions. 
Simulations were performed on computational resources provided by the University of Chicago Research Computing Center.
L.O. acknowledges the support of the Multimessenger Plasma Physics Center (MPPC), NSF grant PHY2206607. 
D.C. was partially supported by NASA through grants 80NSSC20K1273 and 80NSSC18K1218 and NSF through grants AST-1909778, PHY-2010240, and AST- 2009326.
L.S. was supported by NSF grant PHY2409223. This research was facilitated by Multimessenger Plasma Physics Center (MPPC) NSF grants PHY2206607 and PHY2206609 to A.S. and L.S. The work was supported by a grant from the Simons Foundation (MP-SCMPS-0000147, to L.S. and A.S.). L.S. also acknowledges support from DoE Early Career Award DE-SC0023015.


\end{acknowledgments}
\bibliography{Total}
\bibliographystyle{aasjournalv7}

\appendix

\begin{figure*}[h]
    \begin{center}\includegraphics[width=0.94\textwidth, clip=true,trim= 0 55 0 0]{"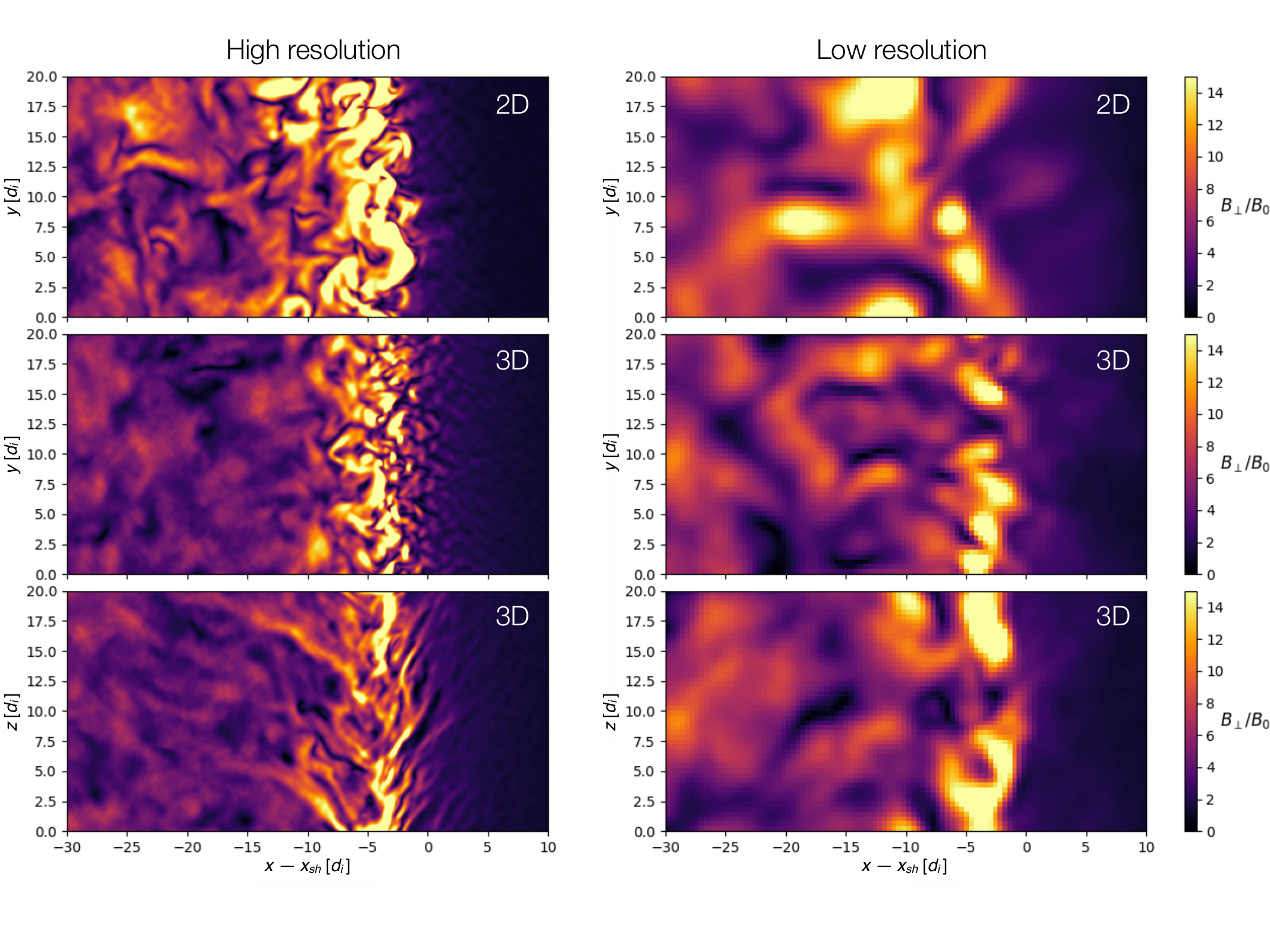"}
    \caption{We report the $B_{\perp}$ profiles for low- and high-resolution simulations with $M_A = 30$, comparing 2D (top panel), a slice in the $x–y$ plane run of the 3D run (middle panel) and a slice in the $x–z$ plane of the 3D run (bottom panel).
    } 
    \label{Fig:alla_fields}
    \end{center}
\end{figure*}

In Figure~\ref{Fig:alla_fields}, we report the $B_{\perp}$ profiles for low- and high-resolution simulations with $M_A = 30$, comparing 2D (top panel), a slice in the $x$–$y$ plane run of the 3D run (middle panel) and a slice in the $x–z$ plane of the 3D run (bottom panel).
In 2D, the downstream region of the shock appears to be composed of extended structures—effectively ``walls"—of strong $B_\perp$. In contrast, in 3D simulations, such structures are absent and there are holes of small $B_{\perp}$. 
For instance, looking at the bottom panels of Figure~\ref{Fig:alla_fields}, where a slice of the $x–z$ plane is shown, in high-resolution 3D simulations, the shock transition is dominated by tilted magnetic tubes, separated by regions of very weak $B_\perp$. For low resolution simulations, this low $B_{\perp}$ regions are even larger. For example, see the region between 7.5 and 12.5 $d_i$ along $z$ in bottom right panel of Figure~\ref{Fig:alla_fields}.

\end{document}